\documentclass[12pt]{article}
\usepackage{graphicx}

\newcommand\pubnumber{SNSN-323-63}
\newcommand\pubdate{\today}

\def\cycl{Cyclotron Institute, Texas A\&M University\\
College Station, TX, 77843-3366 U.S.A.}
\def\support{\footnote{This material is based upon work
    supported by the U.S. Department of Energy, Office of Science, Office of Nuclear Physics, under Award Number DE-FG03-93ER40773, 
    and by the Welch Foundation under Grant No.~A-1397.}}

\def\Title#1{\begin{center} {\Large #1 } \end{center}}
\def\Author#1{\begin{center}{ \sc #1} \end{center}}
\def\Address#1{\begin{center}{ \it #1} \end{center}}

\newcommand\pubblock{\rightline{\begin{tabular}{l} \pubnumber\\
         \pubdate  \end{tabular}}}
\newenvironment{Abstract}{\begin{quotation}  }{\end{quotation}}
\newenvironment{Presented}{\begin{quotation} \begin{center} 
             PRESENTED AT\end{center}\bigskip 
      \begin{center}\begin{large}}{\end{large}\end{center} \end{quotation}}

\def\beq{\begin{equation}}
\def\eeq#1{\label{#1}\end{equation}}
\def\eeqn{\end{equation}}

\def\beqa{\begin{eqnarray}}
\def\eeqa#1{\label{#1}\end{eqnarray}}
\def\eeqan{\end{eqnarray}}

\let\bar=\overbar

\def\Dslash{\not{\hbox{\kern-4pt $D$}}}
\def\dslash{\not{\hbox{\kern-2pt $\del$}}}

\def\ee{e^+e^-}

\def\GF{G_F}

\def\msb{{\bar{\ssstyle M \kern -1pt S}}}

\begin{document}

\def\QEC{Q_{\mbox{\tiny EC}}}
\def\GV{G_{\mbox{\tiny V}}}
\def\Vud{V_{\mbox{\scriptsize ud}}}
\def\Vus{V_{\mbox{\scriptsize us}}}
\def\Vub{V_{\mbox{\scriptsize ub}}}
\def\F{{\cal F}}
\def\DRV{\Delta_{\mbox{\tiny R}}^{\mbox{\tiny V}}}
\def\GF{G_{\mbox{\tiny F}}}
\def\be {\begin{equation}}
\def\ee {\end{equation}}
\def\bea {\begin{eqnarray}}
\def\eea {\end{eqnarray}}

\begin{titlepage}
\pubblock

\vfill
\Title{The current evaluation of $\Vud$}
\vfill
\Author{ J.C. Hardy\support and I.S. Towner$^1$}
\Address{\cycl}
\vfill
\begin{Abstract}
The $\Vud$ element of the Cabibbo-Kobayashi-Maskawa matrix can be determined from several different experimental approaches: 
either $0^+$$\rightarrow$$0^+$ superallowed nuclear $\beta$ decays, neutron decay, nuclear mirror decays, or pion $\beta$ decay.
Currently all give consistent results but, because the nuclear superallowed value has an uncertainty at least a factor of seven
less than all other results, it dominates the result.  A new survey of world superallowed-decay data establishes the $\F t$ values
of 14 separate superallowed transitions to a precision of order 0.1\% or better; and all 14 are statistically consistent with one
another.  This very robust data set yields the result $\Vud$ = 0.97417(21), the value we recommend.   
\end{Abstract}
\vfill
\begin{Presented}
CIPANP2015 \\
Twelfth Conference on the Intersections of Particle and Nuclear Physics  \\ [2mm]
Vail, CO, U.S.A., May 19-24, 2015
\end{Presented}
\vfill
\end{titlepage}
\def\thefootnote{\fnsymbol{footnote}}
\setcounter{footnote}{0}

\section{Superallowed nuclear beta decay}

Beta decay between nuclear analog states of spin-parity, $J^{\pi} = 0^+$, and isospin, $T = 1$, has a unique simplicity: It is a pure
vector transition and is nearly independent of the nuclear structure of the parent and daughter states.  Such transitions are called
``superallowed."  Their measured strength -- expressed as an ``$ft$ value" -- can be related directly to the vector coupling constant
for semi-leptonic decays, $\GV$, with the intervention of only a few small ($\sim$1\%) calculated terms to account for radiative and
nuclear-structure-dependent effects. Once $\GV$ has been determined in this way, it is only another short step to obtain a value for
$\Vud$, the up-down mixing element of the Cabibbo-Kobayashi-Maskawa (CKM) matrix. 

The $ft$ value of any $\beta$ transition is simply the product of the phase-space factor, $f$, and the partial half-life of the transition,
$t$.  It depends on three measured quantities: the total transition energy, $\QEC$, the half-life, $t_{1/2}$, of the parent state, and the
branching ratio, $R$, for the particular transition of interest. The $\QEC$ value is required to determine $f$, while the half-life and
branching ratio combine to yield the partial half-life.

In dealing with superallowed decays, it is convenient to combine some of the small correction terms with the measured $ft$-value and
define a corrected $\F t$-value. Thus, we write \cite{HT15}
\be
\F t \equiv ft (1 + \delta_{\mbox{\tiny R}}^{\prime}) (1 + \delta_{\mbox{\tiny NS}} - \delta_{\mbox{\tiny C}} ) = \frac{K}{2 \GV^2 
(1 + \DRV )}~,
\label{Ftconst}
\ee
where $K = 8120.2776(9) \times 10^{-10}$ GeV$^{-4}$s, $\delta_{\mbox{\tiny C}}$ is the isospin-symmetry-breaking correction and $\DRV$
is the transition-independent part of the radiative correction. The terms $\delta_{\mbox{\tiny R}}^{\prime}$ and $\delta_{\mbox{\tiny NS}}$
constitute the transition-dependent part of the radiative correction, the former being a function only of the electron's energy and the $Z$
of the daughter nucleus, while the latter, like $\delta_{\mbox{\tiny C}}$, depends in its evaluation on nuclear structure. From this equation,
it can be seen that a measurement of any one superallowed transition establishes a value for $\GV$. The measurement of several tests the
Conserved Vector Current (CVC) hypothesis that $\GV$ is not renormalized in the nuclear medium. If indeed $\GV$ is constant -- i.e. all the
$\F t$-values are the same -- then an average value for $\GV$ can be determined and $\Vud$ obtained from the relation $\Vud = \GV/\GF$, where
$\GF$ is the well known \cite{PDG14,Ti13} weak-interaction constant for purely leptonic muon decay.

It is important to note that if, instead, the $\F t$ values show a significantly non-statistical inconsistency, 
one to the other, then the remaining steps cannot be taken since inconsistency would demonstrate that the correction
terms were not correct or, less likely, that CVC had been violated.  Without consistency, there is no coupling
``constant" and there can be no justification for extracting a value for $\Vud$.

\begin{figure}[t]
\centering
\includegraphics[width=13.6 cm]{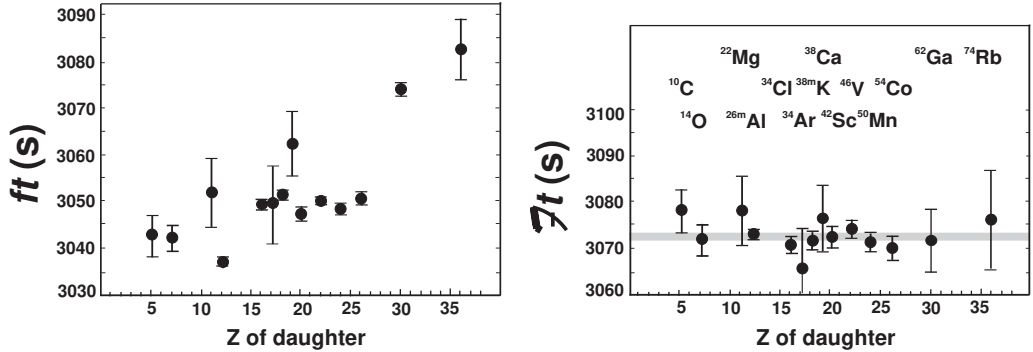}
  \caption{Results from the 2015 survey \protect\cite{HT15}: uncorrected $ft$ values for the 14 best
known superallowed decays on the left; the same results but incorporating the $\delta_{\mbox{\tiny R}}^{\prime}$,
$\delta_{\mbox{\tiny C}}$ and $\delta_{\mbox{\tiny NS}}$ correction terms on the right.  The grey band in the right panel is the average $\F$$t$
value and its uncertainty.}
\label{fig1}
\end{figure}

Early in 2015, we published \cite{HT15} a new critical survey of all half-life, decay-energy and branching-ratio
measurements related to 20 superallowed $0^+$$\rightarrow 0^+$ $\beta$ decays.  Included were 222 individual
measurements of comparable  precision obtained from 177 published references.  We obtained world-average $ft$
values for each of the 18 transitions that had a complete set of data, and then applied radiative and
isospin-symmetry-breaking corrections to extract corrected $\F t$ values.  A total of 14 of these $\F t$ values
have a precision of order 0.1\% or better; their uncorrected $ft$ values and corrected $\F t$ values are shown
in Fig.~\ref{fig1}.

It is immediately evident from the figure that the $\F$$t$ values are all consistent with one another from $A$=10
to $A$=74.  This simultaneously confirms the CVC expectation of a constant value for $\GV$ and demonstrates the absence
of any significant scalar current, which would introduce an upward or downward curve into the $\F$$t$-value locus at low
$Z$ \cite{HT15}.  It also goes a long way towards validating the particular set of calculated transition-dependent
corrections that were used in the analysis. These calculations of $\delta_{\mbox{\tiny C}}$ and $\delta_{\mbox{\tiny NS}}$
were an updated version of those presented in Ref.~\cite{TH08} and employed the best available shell-model wave functions,
which in each case had been based on a wide range of spectroscopic data for nuclei in the same mass region.  They were
further tuned to agree with measured binding energies, charge radii and coefficients of the isobaric multiplet mass
equation for the specific states involved.  This means that the origins of these correction terms are completely
independent of the superallowed decay data, so consistency in the corrected $\F$$t$ values gives powerful support to the
calculated corrections used in the derivation of those $\F$$t$ values.  We will return later to the question of alternative
calculations for these correction terms.  

With a mutually consistent set of $\F$$t$ values, one is then justified in proceeding to determine the value of $\GV$ and, 
from it, $\Vud$.  The result we obtained from the new survey is
\bea
|\Vud| = 0.97417(21)~~~~~~~~~~~~~~~~~~~~~~~~~~~~~~~  \mbox{[nuclear superallowed]}. \nonumber 
\eea

\section{Other methods for determining $\Vud$}

Neutron $\beta$ decay is the simplest $\beta$ decay to involve both the vector and axial-vector weak interactions.  It is
an attractive option for determining $\Vud$ since its analysis does not require the application of corrections for
isospin-symmetry-breaking, $\delta_{\mbox{\tiny C}}$, or for nuclear-structure-dependent radiative effects,
$\delta_{\mbox{\tiny NS}}$.  However, it has the distinct disadvantage that it requires a difficult correlation measurement
in order to separate the vector-current contribution to its decay from the axial-vector one.  Not only that, but neutrons are
inherently more difficult to handle and contain than nuclei.

Since the $\QEC$ value and the branching ratio for neutron $\beta$ decay are very well known, the crucial measurements required
to determine $\Vud$ are its mean-life and a decay correlation -- usually selected to be the $\beta$ asymmetry from the decay of
polarized neutrons.  World data for both these quantities are not statistically consistent among themselves, the normalized
chi-squared ($\chi^2/N$) for the mean-life average being 3.4 and that for the $\beta$ assymmetry being 3.8.  More alarming still
is the fact that the mean-life results from two different measurement techniques appear to be systematically different from one
another.  The average mean-life obtained when the decay products are recorded from a beam of neutrons is 888.1(20)s; while it is
879.5(7)s when neutrons are confined in a ``bottle" and the survivors are counted a known time later.  It is difficult to know
how to deal with such conflicts so we employ two different methods.  With the first, we follow exactly the same procedures as we
do for the superallowed decays, averaging all world data for each parameter and increasing its uncertainty by the square root of
the normalized chi-squared.  For the second we simply assign a range to the mean-life, which encompasses both the conflicting
sets of results.  The results for $\Vud$ are
\bea 
|\Vud| = 0.9754(14)~~~~~~~~~~~~~~~~~~~~~~~~~~~~~~~  \mbox{[neutron average]},
\nonumber \\
0.9707 \leq \Vud \leq 0.9761~~~~~~~~~~~~~~~~~~~~~~~~~~~~~~~  \mbox{[neutron range]}. \nonumber 
\eea

Neutron $\beta$ decay is just a special case of decay between $T=1/2$ mirror nuclei.  Like neutron decay, these nuclear mirror
decays are mixed vector and axial-vector decays; so, in addition to $\QEC$ values, half-lives and branching ratios, they also
require a $\beta$-asymmetry measurement.  Of course, unlike the neutron, these decays as well require the corrections
$\delta_{\mbox{\tiny C}}$ and $\delta_{\mbox{\tiny NS}}$ for small nuclear-structure-dependent effects.  There are five mirror
decays, $^{19}$Ne, $^{21}$Na, $^{29}$P, $^{35}$Ar and $^{37}$K, for which sufficient data are known.  The relevant world data were
first surveyed in 2008 \cite{Se08}, from which a value of $|\Vud|$ was obtained \cite{Na09}.  More data have appeared since and
been incorporated \cite{Sh14} although there has been very little change in the $|\Vud|$ value obtained.  The current result is
\bea 
|\Vud| = 0.9718(17)~~~~~~~~~~~~~~~~~~~~~~~~~~~~~~~  \mbox{[mirror nuclei]}. \nonumber 
\eea

Finally, the rare pion beta decay, $\pi^+ \rightarrow \pi^0 e^+ \nu_e$, which has a branching ratio of $\sim$$10^{-8}$, is one
of the most basic semi-leptonic electroweak processes.  It is a pure vector transition between two spin-zero members of an isospin
triplet and is therefore analogous to the superallowed $0^+$$\rightarrow$$0^+$ decays.  In principle, it can yield a value of $\Vud$
unaffected by nuclear-structure uncertainties.  In practice, the branching ratio is very small and has proved difficult to measure
with sufficient precision.  The most recent, and by far the most precise, measurement of the branching ratio is by the PIBETA group
\cite{Po04}.  This leads to the result \cite{Bl13}
\bea 
|\Vud| = 0.9749(26)~~~~~~~~~~~~~~~~~~~~~~~~~~~~~~~  \mbox{[pion]}. \nonumber
\eea

\begin{figure}[t]
\centering
\includegraphics[width=11 cm]{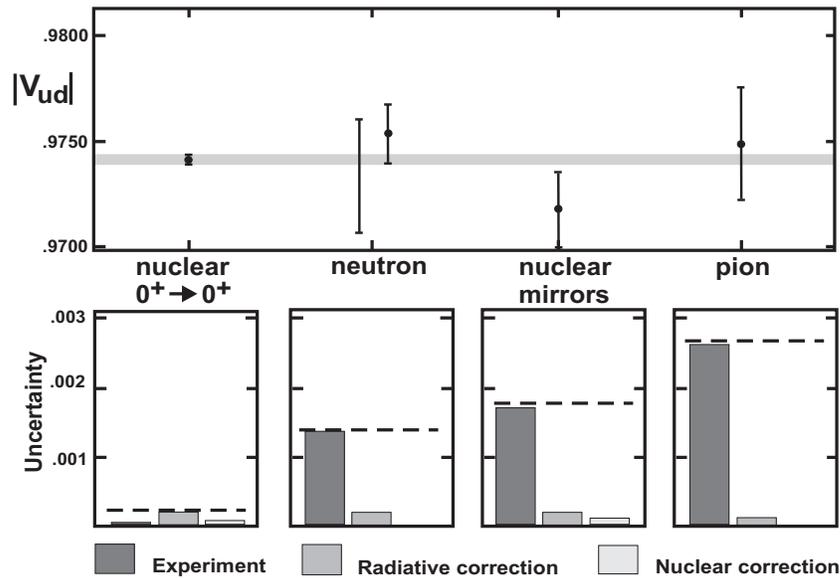}
  \caption{The five values of $|\Vud|$ given in the text are shown in the top panel, the grey band being the average value.  The
four panels at the bottom show the error budgets for the corresponding results with points and error bars at the top.}
\label{fig2}
\end{figure}

\section{Recommended value for $\Vud$}

The five results we have quoted for $|\Vud|$ are plotted in Fig.~\ref{fig2}.  Obviously they are consistent with one another but,
because the nuclear superallowed value has an uncertainty a factor of 7 to 13 smaller than the other results, it dominates the
average.  Furthermore, the more precise of the two neutron results can hardly be considered definitive since it ignores a serious
systematic uncertainty in the data.  Consequently we recommend using the nuclear superallowed result as the best value for $|\Vud|$: i.e.
\be
|\Vud| = 0.97417(21) .
\label{Vud}
\ee

\section{Potential for improvement}

The uncertainty budgets plotted in the bottom panels of Fig.~\ref{fig2} reveal three important facts.  First, experimental uncertainties
dominate in the cases of the less-precisely known neutron, nuclear-mirror and pion $\beta$-decays; while theory makes the largest contribution
to the overall uncertainty for the key $0^+$$\rightarrow 0^+$ decays.  Second, by far the most important theoretical contribution to the latter
is from the radiative correction, principally it turns out \cite{HT15} from $\DRV$, the transition-independent part of the radiative correction.
Finally, the size of the $\DRV$ contribution is the same for all measurement methods; thus we can conclude that no major improvement in the
value of $|\Vud|$ can be achieved in future without improved calculations of $\DRV$.

Unfortunately, experiment can play no role in reducing the $\DRV$ uncertainty.  That must remain a purely theoretical challenge.  The impact of
any improvement would be immediate though: If the $\DRV$ uncertainty were to be cut in half, the $|\Vud|$ uncertainty would be reduced by 30\%.

In the meantime, some small improvement in the $|\Vud|$ uncertainty can still be made with the help of nuclear experiments.  These experiments
can contribute to improving the nuclear-structure-dependent corrections ($\delta_{\mbox{\tiny C}}$ - $\delta_{\mbox{\tiny NS}}$), which produce
the second largest component of the $|\Vud|$ uncertainty budget for the $0^+$$\rightarrow 0^+$ decays (see Fig.~\ref{fig2}).  In the past few
years, a number of different groups have published $\delta_C$ values from calculations based upon a variety of different model approaches.
Typically each calculation covers only a subset of the measured transitions but the subsets are not the same from calculation to calculation and,
where overlap does exist, the results are not notably consistent with one another.  This diversity of results has prompted us to develop a test
\cite{TH10} to assess the quality of each calculated set of corrections and determine its relative merit.  The test is based on the premise that
the CVC hypothesis is valid and thus the corrected $\F t$ values for all measured transitions should be statistically consistent with one another
($i.e.$ with $\chi^2/N$$\sim$1).  

This test has already contributed to reducing the uncertainty in $\delta_{\mbox{\tiny C}}$.  As part of our recent survey \cite{HT15}, we applied
the test to all sets of calculations that cover at least half the number of well-measured superallowed transitions.  The resultant $\chi^2/N$ values
for the various calculations spanned a wide range, with only a single set \cite{TH08} yielding a value near one.  In this way, we identified that set
as the one to use in our ultimate analysis of the experimental data (see Fig.\,\ref{fig1}).  No allowance for systematic differences between
otherwise acceptable calculations was required since no other set passed the acceptability test.

There is a second test that can be expected to refine the selection process for $\delta_C$ calculations even further.  It involves the measurement
of mirror pairs of superallowed transitions, which has only just become possible, with the first case --- $^{38}$Ca $\rightarrow$ $^{38m}$K and
$^{38m}$K $\rightarrow$ $^{38}$Ar --- having appeared very recently \cite{Pa14, Pa15}.  This test also depends on the expected constancy of $\F$$t$
values, but in this instance it applies to the two members of a mirror pair of $0^+$$\rightarrow 0^+$ transitions.  Considering current
capabilities for producing superallowed $T_Z$\,=\,-1 parent nuclei in sufficient quantity for a high-statistics measurement, we conclude that
there are three mirror pairs in addition to the one at $A$=38 that can be completed in the immediate future.  These are 
$^{26}$Si\,$\rightarrow$$^{26m}$Al and $^{26m}$Al\,$\rightarrow$$^{26}$Mg; $^{34}$Ar\,$\rightarrow$$^{34}$Cl and $^{34}$Cl\,$\rightarrow$$^{34}$S;
and $^{42}$Ti\,$\rightarrow$$^{42}$Sc and $^{42}$Sc\,$\rightarrow$$^{42}$Ca.

These tests have already played a role in reducing the $|\Vud|$ uncertainty.  With improved measurement precision on the already known $ft$ values,
together with the addition of new mirror pairs of transitions, some modest further improvement can be expected.  However, ultimately these will
only have a significant impact on the $|\Vud|$ result after meaningful improvements have been made in the calculation of $\DRV$.

\section{Note on $|\Vus|$ and the CKM unitarity test} 
  
The standard model does not prescribe the individual elements of the CKM matrix -- they must be determined experimently -- but absolutely fundamental
to the model is the requirement that the matrix be unitary.  To date, the most demanding test of CKM unitarity comes from the sum of squares of
the top-row elements, $|\Vud|^2 + |\Vus|^2 + |\Vub|^2$, which should equal exactly one.  Combining our value for $|\Vud|$ in Eq.~(\ref{Vud}) with the
values of $|\Vus|$ and $|\Vub|$ recommended by the Particle Data Group (PDG) \cite{PDG14}, the top-row sum yields the result 0.99978(55), in excellent
agreement with unitarity.

Unfortunately this cannot be the last word since the PDG evaluation does not include recent results from the most recent lattice calculations, which are
used to extract $|\Vus|$ from semileptonic kaon decays ($K \rightarrow \pi \ell \nu_{\ell}$), and $|\Vus|/|\Vud|$ from the ratio of the pure leptonic decay
of the kaon ($K^{\pm} \rightarrow \mu^{\pm} \nu$) to to that of the pion ($\pi^{\pm} \rightarrow \mu^{\pm} \nu$).  In the past, the results for $|\Vus|$
and $|\Vus|/|\Vud|$ have formed a consistent set with the result for $|\Vud|$.  As the quoted uncertainties on the lattice calculations have been reduced,
however, some tension has appeared, with the combination of results for $|\Vud|$ and $|\Vus|/|\Vud|$ continuing to yield excellent agreement with unitarity
but the combination of $|\Vud|$ and $|\Vus|$ being low by two standard deviations.  This is not a cause for serious concern, but the inconsistency between
the two kaon-decay approaches will need to be resolved in future.

This subject is discussed in more detail in section VB of Ref.~\cite{HT15}.

\end{document}